\documentclass{emulateapj}
\usepackage{apjfonts}

\newcommand\beq{\begin{equation}}
\newcommand\eeq{\end{equation}}

\slugcomment{ApJL, in press}
 
\shorttitle{Flares in Gamma-Ray Bursts}
\shortauthors{Perna, Armitage \& Zhang}

\begin{document}

\title{Flares in long and short Gamma-Ray Bursts: a common origin in a hyperaccreting accretion disk}

\author{Rosalba Perna\altaffilmark{1}, Philip J. Armitage\altaffilmark{1} 
and Bing Zhang\altaffilmark{2}}

\affil{$^1$JILA \& Department of Astrophysical and Planetary 
Sciences, University of Colorado, 440 UCB, Boulder, CO80309-0440; rosalba@jilau1.colorado.edu, 
pja@jilau1.colorado.edu}
\affil{$^2$Department of Physics, University of Nevada, 4505 Maryland Parkway, 
Las Vegas, NV 89154-4002; bzhang@physics.unlv.edu}

\begin{abstract}
Early-time X-ray observations of GRBs with
the {\em Swift} satellite have revealed a more complicated
phenomenology than was known before. In particular, the presence of
flaring activity on a wide range of time scales probably requires late-time
energy production within the GRB engine. Since the flaring activity is 
observed in both long and short GRBs, its origin must be within what 
is in common for the two likely progenitors of the two classes of bursts: 
a hyperaccreting accretion disk around a black hole of a few solar masses. 
Here, we show that some of the observational properties of the flares, 
such as the duration-time scale correlation, and the duration-peak luminosity
anticorrelation displayed by most flares within a given burst, are
qualitatively consistent with viscous disk evolution, provided that 
the disk at large radii either fragments or otherwise suffers large 
amplitude variability. We discuss the physical conditions in the outer parts 
of the disk, and conclude that gravitational instability, possibly followed by
fragmentation, is the most likely candidate for this variability.
\end{abstract}


\section{Introduction}
The launch of the {\em Swift} satellite has opened a new era of GRB
studies. In addition to fulfilling pre-launch expectations by discovering the
afterglows of short GRBs (Gehrels et al. 2005; Fox et al. 2005; Covino
et al. 2005) and detecting GRBs at very high redshift (Haislip et
al. 2005; Cusumano et al. 2005; Tagliaferri et al. 2005; Kawai et
al. 2005), {\em Swift} has also revealed a new, unexpected phenomenology.
{\em Swift} XRT observations have shown that the early X-ray afterglow
lightcurves of nearly a half of {\em Swift} bursts harbor erratic X-ray
flares (Burrows et al. 2005; Nousek et al. 2005; O'Brien et
al. 2005). Although in some bursts there is only one distinct flare
(e.g. GRB 050406, Romano et al. 2005), in other cases there are several flares in each
burst (O'Brien et al. 2005). In particular, the X-ray afterglow
lightcurve of the $z=6.3$ GRB 050904 shows erratic variability
with several distinct flares (Cusumano et al. 2005; Watson et
al. 2005). 

Understanding the origin of the flares is of great theoretical
interest, since they trace the activity of the GRB engine. 
Although the existence of energy injection subsequent to the main 
GRB phase has been used to argue in favor of magnetic rather than 
neutrino jet launching (Fan, Zhang \& Proga 2005), the origin of 
re-energizations in the first place remains unclear.  
For the case of long GRBs, King et al. (2005) suggested that the X-ray
flares could be produced from the fragmentation of the collapsing
stellar core in a modified hypernova scenario. The fragment subsequently 
merges with the main compact object formed in the collapse, releasing 
extra energy. In this two-stage collapse model, the time delay between 
the burst and the flare reflects the gravitational radiation time scale 
for the orbiting fragment to be dragged in. For the case of short GRBs, 
MacFadyen, Ramirez-Ruiz \& Zhang (2005) suggested that the flares could be the result 
of the interaction between the GRB outflow and a non-compact stellar 
companion in a model in which short GRBs result from the collapse 
of a rapidly rotating neutron star in a close 
binary system. 

In this {\em Letter} we suggest a new interpretation of the origin of
the observed X-ray flares. Our model is motivated by the fact
that the flares are observed in both long and short GRBs, which 
are likely to be associated with different types of progenitors:
namely collapsars for the long GRBs (Stanek et al 2003; Hjorth et al. 2003) 
and mergers of compact objects for the short ones (Gehrels et al 2005;
Bloom et al. 2005; Fox et al. 2005; Villasenor et al. 2005; Barthelmy
et al. 2005; Berger et al. 2005). Unless the similarities between these 
classes of events are coincidental, the flares are likely to have something
to do with what is in common between a GRB triggered by a collapsar
and a GRB triggered by a binary merger: an accretion disk rapidly
accreting onto a black hole (BH). In the following we argue that 
the observational properties of the flares are generically consistent 
with some kind of large amplitude instability occurring in the outer 
part of the accretion disk. The origin of such an instability is 
speculative, but we suggest gravitational instability as one 
possibility.

\section{Observational properties of the flares}
Several observed properties of the flares constrain potential theoretical 
explanations. The flares typically rise and fall rapidly, with the typical rising
and falling time scales usually much shorter than the epoch when the flare
occurs. Their arrival time does not appear to correlate with the burst 
duration $T_{90}$. The existence of flares is difficult to interpret within the 
framework of the external shock model, and is consistent with a late internal
shock origin, which requires late central engine activity after the
prompt gamma-ray emission phase is over (Burrows et al. 2005; Zhang et
al. 2005). The internal shock model also demands a much smaller energy
budget than the external shock model (Zhang et al. 2005).
Given the above, we assume henceforth that the flares
directly reflect the activity of the GRB engine.  

Another qualitative feature is that within a particular burst, the durations
of the flares are typically positively correlated with the times when
the flares occur. This is evident from the fact that the widths of the
flare pulses are more or less similar in logarithmic lightcurves
(O'Brien et al. 2005; Cusumano et al. 2005), and that the quantity
$\delta t/t_{peak}$ is essentially constant for different flares
(e.g. Godet et al. 2005). This is particularly clear for the long GRB 050502B
(Falcone et al. 2005) and the short-hard GRB 050724 (Barthelmy et
al. 2005). In both cases, there is an early flare (several 100s for
GRB 050502B and several 10s for GRB 050724) whose duration is of the
order of the peak time itself, and there is also a very late flare at
several times $10^4$s, with a duration of the same order. The luminosity 
of the flares appears to scale with the total energy budget of the 
prompt emission, and hence is smaller for the short burst. Apart from 
this overall normalization, there are no clear differences between flare 
properties in long and short bursts, although there is only one short 
burst in which several flares have been observed. 

Finally, the peak luminosity of the flares is typically
negatively correlated with their arrival time: later flares tend to be 
less bright than earlier ones. This is particularly evident for the
case of GRB 050904 (Watson et al. 2005). 
The total fluence of the flares also tends to decrease with their arrival time,
though to a lesser extent, given their longer duration.

\section{A model for flares in long and short GRBs}
The ultimate source of power in GRBs is believed to be 
black hole accretion\footnote{We note that in scenarios in which 
the role of the accretion flow is primarily to tap into the spin energy 
of the black hole -- for example via the Blandford-Znajek (1977) mechanism -- 
then the net magnetic flux carried inward by the flow is at least as 
important as the mass accretion rate itself.}. Whether the accreting material 
is provided by the envelope of a collapsing star as in the collapsar
model (MacFadyen \& Woosley 1999), or by the debris of a tidally
disrupted companion in binary merger scenarios (Eichler et
al. 1989), the outcome is a disk accreting at very high rates. Within 
the internal shock scenario (e.g. Kobayashi et
al. 1997) variability in GRB light curves
closely reflects the underlying variability of the inner engine. 

The basic physical conditions within GRB disks can be derived 
assuming steady-state conditions 
(e.g. Popham, Woosley \& Fryer 1999; Narayan, Piran \& Kumar 2001; Di 
Matteo, Perna \& Narayan 2002). These studies show that at the high accretion 
rates ($\dot{M}\sim 1M_\odot$ s$^{-1}$) needed to provide the GRB luminosity, 
the disk cannot cool efficiently, and a large fraction of it is advection
dominated. Furthermore, especially in the high temperature inner regions,  
neutrinos play a major role in providing cooling for the disk. Janiuk et
al. (2004) extended these steady-state studies by following the
time-dependent evolution of a neutrino-dominated accretion disk, as its
fuel supply dwindles and its accretion rate drops. Their calculations showed that, after the
material replenishing the disk is exhausted, the accretion rate and resulting 
engine power drop 
rapidly. A rapid drop of the energy deposition rate was also found 
by Setiawan et al (2004) by means of hydrodynamic simulations.   
The observed flares demonstrate that, while this picture
may hold for the main GRB phase, it can fail at late times. We argue below
that the observed flare properties are suggestive of re-energization by blobs 
of material that make their way from a range of initial distances toward the 
accreting black hole. 

The properties of the GRB disk, and in particular the main cooling processes, 
are highly dependent on the mass accretion rate;
for a given accretion rate, they further depend on the radial location 
within the disk (e.g. Narayan et al. 2001; Di Matteo et al. 2002, DPN in the following) 
and on the composition which affects the opacity (Menou, Perna \& Hernquist 2001). 
However, for given disk properties, there exists a typical time scale, the viscous time 
\begin{equation}
t_0(R)=\frac{R^2\Omega_K}{\alpha c^2_s}
\label{eq:t0}
\end{equation}
that sets the typical duration of the accretion phase for a ring of material
initially at a distance $R$ from the accreting object. 
In the above equation, $\Omega_K$ is the Keplerian
velocity of the gas in the disk, $c_s$ the sound speed of the
accreting material, 
and $\alpha$ a parameter characterizing the strength of viscosity 
(Shakura \& Sunyaev 1973). 
The dependence of $t_0$ with $R$ changes depending on the
characteristics of the disk.  
If advection dominates, as found by DPN for accretion rates
$\dot{M}\ga 1 M_\odot$ 
sec$^{-1}$, then the disk scale height $H=c_s/\Omega_K$ is $\sim R$, and the
viscous time scale can be approximated as
\beq 
t_0\sim \frac{1}{\alpha \Omega_K}\sim 5\times 10^{-4}
\alpha^{-1}_{-1}m_3 r^{3/2}\;{\rm sec}\;, 
\label{eq:t0ad}
\eeq
where $m_3=M/(3M_\odot)$, $r=R/R_s$ and $R_s$ is the Schwarzschild radius.  
The viscosity parameter has been written in
units of $\alpha_{-1}\equiv\alpha/0.1$ At a distance of $r\sim 1000$,
the accretion time is $\sim 15\,m_3\alpha_{-1}^{-1}$ sec.

Evidently, the viscous time scales associated with the very high $\dot{M}$, 
advection-dominated flow, are short compared to some of the observed 
flare time scales at any reasonable radii. However, at large radii and / or 
late times, the disk will have a smaller accretion rate. 
As the accretion rate decreases, the fraction of advected energy 
also decreases, and the disk cools more efficiently, implying $H<R$. 
A smaller $H/R$ ratio increases the accretion time $t_0$ by a factor 
of $(H/R)^{-2}$. This can be substantial. 
In the limit of a standard Shakura-Sunyaev disk (which will not, we 
emphasize, be attained during the time interval of interest here), 
$H/R$ is of the order of $10^{-2}$ in the region at $r \sim 10^3$ that 
is dominated by gas pressure and electron scattering opacity. If we assume 
that later-time flares arise from significantly depleted disks, whose 
accretion rate is not high enough to make the disk
substantially advection-dominated, then accretion times of $10^4$ sec 
from material at $r > 10^3$  are not unreasonable. Depending 
upon the conditions then, the accretion time scales from fragments of 
material originally in the outer parts of the disk can vary between tens 
of seconds to several thousands of seconds, with the longer
time scales obviously deriving from the outermost rings of material.

How the derived accretion time scale relates to the observable delays
between the initial event and subsequent flares depends upon the state
of the disk material. If the disk develops a ring-like structure, but
otherwise remains continuous, then the arrival time of a ring of
material initially at a distance $R$ is simply of the order of
$t_0(R)$. Once this ring begins to accrete, the duration of the main
(i.e. most intense) accretion phase is also on the order of $\sim
t_0$. After a time $t\sim t_0$, the accretion rate drops abruptly (see
e.g. numerical simulations by Cannizzo et al. 1990).  Therefore, this
scenario predicts that the arrival time of each new flare should
directly correlate with its total duration. The sudden cessation
of the jet power (i.e. accretion power) would also leave imprints in
the lightcurve by means 
of a rapid decay of flux by means of the so-called ``curvature
effect'' (Kumar \& Panaitescu 2000; Zhang et al. 2005). Such rapid
decays have been indeed seen in the post-flare lightcurves of many
bursts (e.g. Falcone et al. 2005; Barthelmy et al. 2005). 
Moreover, the accretion
rate of a ring of initial mass $M(t_0)$ is of the order of
$\dot{M}(t_0)\propto M(t_0)/(t_0)$ during the time $t_0$.  If the
initial masses of the fragments are not hugely different (or at least
the outermost ones are not substantially more massive than the
innermost ones), then the peak luminosity of
the flares should decline with their arrival times. This again
reflects the general behavior seen in the observed flares. The total
energy (proportional to the fluence) released in each flare is
proportional to the total mass of the fragment that caused that episode
of activity.

Alternatively, it is possible that the disk -- in its outer 
regions -- does not remain a continuous fluid but rather fragments into one or 
more bound objects. This does not alter the generic behavior discussed above, 
but does alter the expected time scales. In particular, if most of the mass at 
large radii is bound up in fragments, then the time scale for those 
objects to be dragged in to the black hole via viscous effects 
is lengthened by a factor of the order of $(M_{\rm frag} / M_{\rm disk}) > 1$, 
where $M_{\rm disk}$ is the exterior disk mass (Syer \& Clarke 1995). This 
could permit substantially longer time scales for any given radius. 
Furthermore, since in a fragmentation scenario new fuel for the 
central engine would be provided via tidal disruption of the 
fragments (at a radius smaller than the initial fragmentation 
radius), the duration of bursts of accretion would 
be shortened. We might expect to observe shorter flares, separated 
by relatively longer intervals between flares.

\section{Disk fragmentation by gravitational instabilities}
The observed characteristics of the X-ray flares
seen in several GRB light curves are consistent with their being 
produced by large changes in the inner accretion rate that are 
controlled by the viscous time scale at large distances from the 
black hole. This immediately begs the question of what
are the important processes in the outer disk that drive such 
variability. The picture we start with is
the one developed through numerical studies of progenitor models:
whether as a result of the collapse of a massive star (see MacFadyen
\& Woosley 1999), or as a result of the merger of two compact objects
(e.g. Ruffert \& Janka 2001; Rosswog et al. 2004; Lee et al. 2005), a massive, 
rapidly accreting disk is formed. The physical conditions within these 
disks are fairly similar, irrespective of their origin. In particular, 
while the total energy output of short bursts
is at least a factor of 10 smaller than that of the long ones, the peak
luminosities in the two classes of objects are comparable (Gehrels et al. 2005;
Villasenor et al. 2005; Bartelmy et al. 2005). This
suggests that, while the total mass of the initial accretion disk is
likely to be smaller for the short bursts, the accretion rates during
the prompt phase are comparable for the two classes of bursts and can range
from a tenth of $M_\odot$ s$^{-1}$ or so, to several times $M_\odot$ s$^{-1}$.

Several classes of known disk instabilities could in principle lead to 
large variations in the accretion rate onto the black hole (as noted 
above, the seat for these instabilities must lie in the outer disk 
from time scale arguments). Thermal instability occurs where 
$({\rm d} \ln Q^+ / {\rm d} \ln T)_\Sigma > ({\rm d} \ln Q^- / {\rm d} \ln T)_\Sigma$
where $Q^{-}$ and $Q^{+}$ are the cooling and heating rates and $\Sigma$ is 
the surface density of the disk, while if ${\rm d}\dot{m} / {\rm d}\Sigma < 0$ 
the disk would be viscously unstable. Thermal instability in particular 
is known to lead to very large amplitude outbursts, for example in 
dwarf novae. However, a stability analysis of the GRB disk models by DPN 
suggested that these disks are thermally and viscously stable throughout 
their entire radial extent, so these instabilities do not appear promising 
as explanations for flares. Also disfavored are {\em small-scale} intrinsic 
instabilities, whether associated with MHD turbulence (De Villiers, Hawley \& 
Krolik 2003) or with shock instabilities associated with the photodisintegration 
of $\alpha$ nuclei (MacFadyen \& Woosely 1999). Although variability associated 
with these processes is inevitable -- and likely responsible for the variability 
in the accretion rate necessary to explain the variable luminosity of the GRB
prompt emission -- they do not yield large-scale coherent flares as apparently 
required to match observations.

A better, though still speculative, possibility is gravitational instability, 
especially if that instability results in actual fragmentation of the disk. 
The accretion flow will become gravitationally unstable if 
the Toomre parameter (Toomre 1964),
\begin{equation}
Q_T = { {c_s \kappa} \over {\pi G \Sigma} } < 1,
\end{equation}
where $c_s$ is the sound speed and $\kappa$ the epicyclic frequency. 
This may occur in the outer regions. DPN found that a gravitationally unstable
region ($Q_T \sim 1$) is possible for $R \gtrsim 10^2 R_s$ if the
accretion rate is high enough (of the order of $\dot{M} \gtrsim 3 M_\odot\,
{\rm s}^{-1}$) at those radii.  For accretion flows arising from
collapsars, this requires at a minimum that the specific angular
momentum of infalling matter be $j \gtrsim 2 \times 10^{17} \ {\rm
cm^2\,s^{-1}}$ (for a $3M_\odot$ central object), which is possible 
if the collapse results from a relatively rapidly rotation stellar core. 
Producing a sufficiently massive disk at large 
radii as a consequence of a merger of compact objects is likely to be harder,  
since at least initially the relatively low specific angular 
momentum implies that $Q_{T,{\rm min}} \sim 10$ (Lee et al. 2005).  
This might suggest that fragmentation is more likely in the case of 
GRB's whose progenitors are collapsars, whereas disks from the merger of 
compact object are more likely to remain continuous. However, since 
observationally flares in the two classes of GRBs appear to be similar 
(although, as noted previously, the observational sample for the 
short GRBs is very limited) a common mechanism for both types is 
favored. In fact, binary merger simulations by Setiawan et al. (2004) show that,
for an initial disk mass on the order of a tenth of solar
mass or so, the peak accretion rate can become as high as
$\dot{M}\sim 10M_\odot$~s$^{-1}$ or more. Under these conditions,
gravitational instability could occur at a few tens of Schwarchild's radii (DPN). 
Moreover, three-dimensional simulations of coalescing neutron stars by Ruffert \& 
Janka (2001) show that a substantial mass of material gains angular 
momentum as a consequence of tidal effects and moves to larger radii 
(indeed, in their models, up to $\sim 30$\% of the total mass in the 
disk actually becomes unbound). 
Therefore, although less likely than for the collapsar model, it is possible
that the conditions for the onset of the gravitational instability might
also occur within disks produced by mergers of compact objects. In this 
case, instability seems most likely to occur as an expanding flow reaches 
the larger radii where gravitational instability is inevitable. 

Once a gaseous disk becomes 
gravitationally unstable, two classes of behavior are possible. First,  
the disk may develop a quasi-steady spiral structure -- which acts to 
transport angular momentum outward and mass inward. This mode of 
gravitational instability can drive large amplitude outbursts of 
accretion if the disk mass is sufficiently large (Laughlin et al. 1998; 
Lodato \& Rice 2005; Vorobyov \& Basu 2005).
Second, the disk may 
fragment into bound objects. Fragmentation is inevitable if the 
local cooling time,
\begin{equation}
 t_{\rm cool} < t_{\rm crit} \approx 3 \Omega_K^{-1}\;,
\end{equation}
in which case the maximum stress obtainable from gravitational 
instability (roughly $\alpha_{\rm max} \sim 0.1$) produces 
insufficient heating to offset cooling (Gammie 2001; Rice et al. 2003). 
For a collapsar disk in its early, advection-dominated phase, we expect 
that the cooling time scale, dominated at these radii by neutrino emission, 
will be comparable to the viscous time scale $t_0 \sim \alpha^{-1} \Omega_K^{-1}$.
For $\alpha = 0.1$ the disk is therefore nominally
stable against fragmentation, though not by such a margin that details
of the problem (such as the presence of cooling via
photodisintegration) might not change the answer.
Fragmentation may also be possible, even for longer cooling time scales, 
if infall adds mass to the disk on a time scale shorter than the 
minimum viscous time scale $t_0 \sim (H/R)^{-2} \alpha_{\rm max}^{-1} 
\Omega_K^{-1}$. 

If fragments form, their initial mass $M_{\rm frag} \sim \Sigma (2H)^2$ is 
set by local disk properties. However, such fragments will 
accrete and / or merge rapidly, until their tidal influence 
on the surrounding disk manages to open a gap. This occurs at a mass (Takeuchi et al. 1996)
\begin{equation}
 M_{\rm frag} \simeq \left( {H \over R} \right)^2 \alpha^{1/2} M
\end{equation}
where $M$ is the central object mass. If the disk fragments in the 
early, advection-dominated phase where $H \sim R$, we would 
expect rather massive fragments, whereas fragmentation taking place 
at larger radii and / or later times would give rise to lower mass 
objects. 

\section{Summary}
Early X-ray observations {\em Swift} have revealed a new, 
rich, and unexpected phenomenology of GRBs.
The observation of energetic flares occurring tens of seconds to tens of
thousands of seconds after the initial burst signal the presence of energy 
injection long after the prompt phase is over. Flares are an important diagnostic 
of the GRB engine. In this {\em Letter} we have noted how the observed positive
correlation between the arrival time and duration of the flares, and the
negative correlation of their peak luminosity with duration (which
most flaring episodes within a given burst display), can be
interpreted. Generically, these observations are consistent with a model 
in which the flares are due to blobs of material which
initially circularize at various radii and subsequently evolve
viscously. Since the flares are seen in both long and short GRBs, and
these are likely to be produced by different progenitors, we have suggested that these
fragments of material are likely to be created within the initial, hyperaccreting
accretion disk. In particular, we have noted how, in the outer parts of these disks,
the physical conditions may be suitable for gravitational instability, leading 
either to large amplitude changes in the inner accretion rate or complete 
fragmentation of the disk followed by relatively slow inspiral of the 
fragments toward the black hole.

\acknowledgments 
We thank Dale Frail, Daniel Proga, Luigi Stella and an anonymous referee for 
helpful comments. 
RP acknowledges support from NASA under grant NNG05GH55G, and
from the NSF under grant AST~0507571.
PJA acknowledges support from NASA under grants NAG5-13207, NNG04GL01G and 
NNG05GI92G, and
from the NSF under grant AST~0407040.
BZ acknowledges support from NASA under grants NNG05GB67G, NNG05GH92G,
and NNG05GH91G.

\end{document}